\documentclass[journal]{IEEEtran}
\usepackage{amsmath,amsfonts,subfigure,adjustbox}
\usepackage{algorithmic}
\usepackage{algorithm}
\usepackage{array}
\usepackage[caption=false,font=normalsize,labelfont=sf,textfont=sf]{subfig}
\usepackage{textcomp}
\usepackage{stfloats}
\usepackage{url}
\usepackage{verbatim}
\usepackage{graphicx}
\usepackage{cite}
\usepackage{makecell}
\usepackage{color}
\newtheorem{Remark}{Remark}

\newtheorem{Theorem}{Theorem}
\newtheorem{Lemma}{Lemma}

\begin{document}
\title{Non-Orthogonal Pilot Sequence Design for Multi-Cells Interference Networks}
		
\author{Zhi Gu and Wai Ho MOW, \textit{Senior Member, IEEE}
\thanks{Z. Gu \& W. Mow are with the Department of Electronic and Computer Engineering, Hong Kong University of Science and Technology, Hong Kong, China. (e-mail: eezhigu@ust.hk; eewhmow@ust.hk)}
}
		
\markboth{Journal of \LaTeX\ Class Files,~Vol.~14, No.~8, August~2024}%
{Shell \MakeLowercase{\textit{et al.}}: A Sample Article Using IEEEtran.cls for IEEE Journals}
		

\maketitle

\begin{abstract}
In wireless communications, the performance of non-orthogonal sequence sets significantly affects the level of multi-user interference when the number of users surpasses the sequence length.
The design of non-orthogonal sequences plays a crucial role in both the non-orthogonality of the pilots in multi-cell systems and the signature sequences in overloaded code-division multiple-access (CDMA) systems.
In multi-cell systems, considering the strength disparity between channels originating from the home cell and the neighboring cells, the extended total squared correlation (ETSC) is proposed as a new sequence design criterion, which is defined as the sum of squares of the weighted correlations among sequences.
In this paper, we derive a closed-form expression for the lower bound of ETSC for multi-cell systems with a given sequence length $\tau$, where $\tau \leq K$ and $K$ is the number of users per cell.
This can be regarded as a generalization of the well-known Welch bound (Welch, 1974, IEEE TIT) and the extended Welch bound (Wang et al., 2021, IEEE TWC).
Additionally, from the necessary conditions of the bound, the optimal sequence set can be easily obtained when the interference power factor matrix is positive definite.
On the other hand, to address the lack of sequence generation methods under certain parameter conditions, we propose the ETSC-MM algorithm, which generates sequence sets with low ETSC based on a Majorization-Minimization (MM) optimization framework.
\end{abstract}
		
		
\section{Introduction}
As the network grows denser and the number of users increases significantly, interference control has become one of the major challenges in future wireless communications \cite{zahir2012interference}.
Pilot sequences play a crucial role in communication systems, handling tasks such as channel estimation and signal detection, which makes interference control of pilot sequences especially important \cite{fan1996sequence}.
Due to the limited duration of coherence time and the large number of users, the pilot sequences for all users are typically non-orthogonal.
This phenomenon is known as pilot contamination \cite{jose2011pilot}. This closely resembles the situation in overloaded code-division multiple access (CDMA) systems, where user signature sequences cannot be made mutually orthogonal \cite{viswanath1999optimal}.

System performance requirements for the 3rd Generation Partnership Project (3GPP) Long Term Evolution (LTE) Advanced \cite{ran2008requirements} and New Radio (NR) target significant improvements in cell-edge spectral efficiency and peak transmission rates that can be reached.
In order to achieve these targets, dense frequency reuse of the scarce radio spectrum allocated to the system is needed. However, this exacerbates the issue of pilot contamination, as it now involves addressing both intra-cell and inter-cell interference \cite{hamza2013survey, yang2015interference}.

Intra-cell interference and inter-cell interference greatly limit system performance, especially for users located at the cell edge.
There is extensive research on handling these issues, including methods such as inter-cell interference coordination \cite{hamza2013survey}, massive MIMO spatial diversity \cite{teeti2015impact}, and Bayesian estimation techniques \cite{yin2013coordinated}, etc.
Despite their promising performance, these methods have certain drawbacks, such as leading to more complex resource scheduling, or increasing hardware overhead, or assuming certain channel statistical information.
Therefore, finding non-orthogonal sequence sets with lower interference is a more fundamental solution to reducing pilot contamination based on fundamental principles \cite{wang2020extending}.

In the literature of non-orthogonal sequence design, most studies are based on treating intra-cell and inter-cell interference equally.
In such cases, total squared correlation (TSC) becomes the criterion for measuring non-orthogonal sequence sets, and the overall interference among users is characterized by the Welch bound \cite{viswanath1999optimal}.
The optimal sequence sets are those whose TSC meets the Welch bound, and they are termed Welch-bound-equality (WBE) sequence sets \cite{welch1974lower}.
Fourier matrices \cite{ding2006complex}, Hadamard matrices \cite{li2009minimum}, and difference sets \cite{xia2005achieving} are commonly used tools for constructing WBE sequence sets. There is extensive research on WBE sequence sets \cite{ding2007generic, hu2013new, yan2022construction}, and due to space limitations, not all of these methods are listed here.
Furthermore, considering the different distances between users and base stations, a total weighted square correlation (TWSC) with power coefficients is proposed \cite{cotae2003optimal}.
In essence, TWSC is an extension of TSC where a parameter, the power coefficient, is added for each user. This allows TWSC to better measure the non-orthogonality under different received power conditions compared to TSC.
Similar to the previous Welch bound, the generalized Welch bound serves as the theoretical lower bound for TWSC, and the sequence set that achieves the generalized Welch bound is called the generalized WBE sequence set \cite{cotae2004total}.
Similarly, there is extensive research on WBE sequence sets \cite{cotae2003optimal, cotae2003construction, cotae2003distributive, cotae2004total}, and due to space limitations, not all of these methods are listed here.

It is worth noting that in practical applications, the strength disparity between the channels originating from the home cell and the neighboring cells should be taken into account, i.e., intra-cell and inter-cell interference should be treated differently \cite{wang2020extending}.
Motivated by the above observations, \cite{wang2020extending} focuses on the case of two cells ($J=2$). Its main contributions in sequence design include two points: a) The inter-cell interference power factor $\beta$ is incorporated into the sequence set design, and extended total squared correlation (ETSC) is proposed as the criterion of sequence design based on $\beta$.
Besides, an extended Welch bound is proposed as a modified version of the traditional Welch bound, which is the lower bound of ETSC.
b) An optimization algorithm is proposed to generate sequence sets that can meet the extended Welch bound.
Note that the number of users $K$ in a single cell and the sequence length $\tau$ must satisfy $K\leq \tau \leq 2K$ \footnote{When
$2K\leq \tau$, this case is trivial. The reason is that all the sequences allocated to users are orthogonal. We believe the authors of \cite{wang2020extending} recognized this point, but did not explicitly mention this trivial case.}.

\subsubsection*{Motivation and Contributions}
Following an analysis of the engineering background, we have recognized three key shortcomings in \cite{wang2020extending}:
a) In practical applications, the number of cells to be considered is not limited to two; typically, three or even more cells need to be taken into account \cite{kosta2012interference}.
b) For a single cell, the relationship between the number of users and the sequence length does not necessarily satisfy $K\leq \tau \leq 2K$.
The reason is that the number of users that base stations (BSs) need to serve varies in different scenarios.
Hence, the case where $\tau \leq K$ should also be taken into consideration \cite{zyren2007overview}.
c) The method proposed in \cite{wang2020extending} generates non-unimodular sequences, which leads to a higher Peak-to-Average Power Ratio (PAPR), thereby increasing the cost of electronic components such as power amplifiers and digital-to-analog converters \cite{litsyn2007peak}.
From a technical perspective, the core reason for these shortcomings is that the sequence generation method in \cite{wang2020extending} heavily relies on the extended Welch bound.
In \cite{wang2020extending}, the authors first provide the extended Welch bound and the condition of achieving this bound, and then find sequences that satisfy the condition based on optimization algorithms.
On the other hand, due to the conditions for equality in the Cauchy-Schwarz inequality, the extended Welch bound can only support two cells and satisfy a restriction of $K\leq \tau \leq 2K$.
Therefore, designing $J(\geq 2)$ cells with $K$ being independent of $\tau$ has become a challenge in engineering.
Moreover, since the method in \cite{wang2020extending} generates sequences by satisfying the conditions for achieving the extended Welch bound, generating unimodular sequences is a challenging problem for this approach.

In our work, we are committed to addressing the above three shortcomings.
We consider an interference system with $J(\geq 2)$ cells. Inspired by \cite{wang2020extending}, we introduce an interference power factor matrix $\mathbf{B}$, which can be used to describe the interference levels between any two different cells.
Next, we generalize ETSC such that it can serve as the criterion for sequence design for $J(\geq 2)$ cells.
Finally, we present a new extended Welch bound, which serves as a complement to the previous extended Welch bounds.
And the necessary condition for achieving the bound is proposed.
Besides, an optimization algorithm is also proposed, which can generate a sequence set with low ETSC.
Note that the sequence sets generated by the above two methods can both (almost) meet their respective theoretical bounds.
The main contributions of this paper can be summarized as follows:
\begin{itemize}
\item Inspired by the Inner Product Theorem \cite{welch1974lower}, we present the Cross-inner Product Theorem, which serves as a cross form of the previous Inner Product Theorem.
Furthermore, based on the Cross-inner Product Theorem, we propose a new extended Welch bound for the case of $K\geq\tau$, $J\in \mathbb{N}^+$ and $\mathbf{B}$ is a positive definite matrix.
Additionally, we have provided a necessary condition to achieve the bound: each cell should be assigned a WBE sequence set.
Compared to the previous extended Welch bound \cite{wang2020extending}, which applies to the case of $K \leq \tau$ and $J=2$, our new bound provides a complementary perspective.
\item Based on the Majorization-Minimization (MM) optimization framework \cite{hunter2004tutorial}, we propose the ETSC-MM algorithm to generate sequence sets with low ETSC.
Unlike the method in reference \cite{wang2020extending}, the ETSC-MM algorithm does not need to find the conditions that achieve the extended Welch bound. Instead, it directly uses ETSC as the objective function and implements optimization through two MM operations.
Therefore, the ETSC-MM algorithm can work independently of the extended Welch bound, i.e., it can address the pilot sequence generation problem for $J$ cells.
Moreover, it imposes no limit on the sequence length and the positive definiteness of matrix $\mathbf{B}$.
In simple terms, the ETSC-MM algorithm effectively addresses the shortcomings of the theoretical method based on WBE sequence set and the numerical algorithm in \cite{wang2020extending}.
\end{itemize}


\subsubsection*{Organization}
The remaining sections of the paper are organized as follows.
In Section II, the problem formulations are presented.
In Section III, we propose a new extended Welch bound.
In Section IV, we first give a brief review of the MM framework and then the ETSC-MM algorithm is derived, followed by the complexity and convergence analysis.
Finally, Section V presents some numerical examples and simulations.
The conclusions are given in Section VI.

\section{Preliminaries}\label{sec2}
The following notations will be used throughout this paper.
\begin{itemize}
\item $\mathbf{X}^*, \mathbf{X}^T$ and $\mathbf{X}^H$ denote the complex conjugate, the transpose and the conjugate transpose of matrix $\mathbf{X}$, respectively;
\item $\langle\mathbf{a},\mathbf{b}\rangle$ denotes the inner-product between two complex valued sequences $\mathbf{a}=[a[0],a[1],\ldots,a[N-1]]^T$, $\mathbf{b}=[b[0],b[1],\ldots,b[N-1]]^T$, i.e., $\langle\mathbf{a},\mathbf{b}\rangle= \sum_{k=0}^{N-1}a[k]b^*[k]$, where $N$ is the sequence length of $\mathbf{a}$ (and $\mathbf{b}$);
\item $[\mathbf{a}]_{k}$ denotes the $k$-th element of sequence $\mathbf{a}$;
item $[\mathbf{X}]_{i,j}$ denotes the $i$-th row $j$-th column element of matrix $\mathbf{X}$;
\item $\zeta_N$ denotes the $N$-th complex roots of unity, i.e., $\zeta_N=e^{2\pi i/N}$;
\item $\mathbf{F}_N$ denotes the Fourier matrix of size $N$, i.e., $[\mathbf{F}_N]_{i,j}=\frac{1}{\sqrt{N}}\zeta_N^{-(i-1)(j-1)}$;
\item $\mathbf{a}\odot\mathbf{b}$ denotes element-wise multiplication, i.e., $[\mathbf{a}\odot\mathbf{b}]_k=a[k]b[k]$;
\item $\mathbb{E}(\cdot)$ denotes the expected value of a random variable;
\item $\varphi(\mathbf{a})$ denotes the phase of each of the elements of $\mathbf{a}$.
\end{itemize}

\subsection{System Model}
\begin{figure}[ht]
  \centering
  \includegraphics[width=8cm]{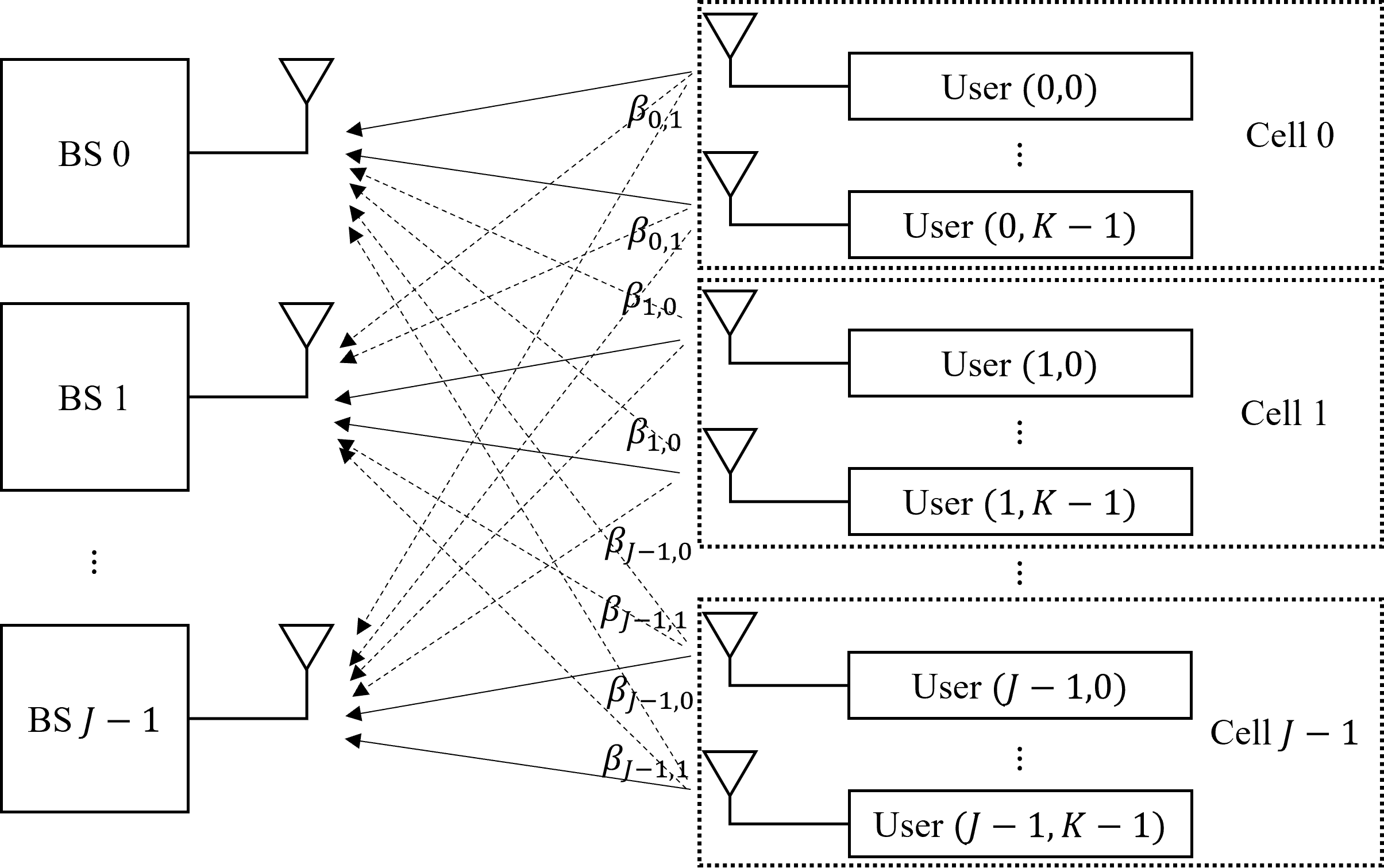}
  \caption{$J$-cell multi-user interference network.}\label{fig_modle.png}
\end{figure}

As shown in Fig. \ref{fig_modle.png}, we consider the uplink of a $J$-cell system. In each cell, there are $K$ single-antenna users communicating with the BS.
We employ a Wyner-type system model \cite{wyner1994shannon, wang2020extending}, which is usually used when investigating cellular networks due to its simplicity and analytical tractability.
In this model, an inter-cell interference power factor matrix $\mathbf{B}$ is assumed to represent the intensity of the inter-cell interference, which is given by
\begin{equation}\label{equation_B}
    \mathbf{B}=
\left[
  \begin{array}{cccc}
    \beta_{0,0} & \beta_{0,1} & \cdots & \beta_{0,J-1} \\
    \beta_{1,0} & \beta_{1,1} & \cdots & \beta_{1,J-1} \\
    \vdots & \vdots &  & \vdots \\
    \beta_{J-1,0} & \beta_{J-1,1} & \cdots & \beta_{J-1,J-1} \\
  \end{array}
\right],
\end{equation}
where $\beta_{m,n}\in [0,1]$ represents the interference intensity from the users of the $m$-th BS to the users of the $n$-th BS.
Here, we always assume $\beta_{m,m} = 1$, as it dominates the system.
Block fading is assumed where the channel realizations are constant over one coherent block with block length $T$ and independent across blocks \cite{hassibi2003much, wang2020extending}.
The entries of the involved channels are independent and identically distributed (i.i.d.) complex Gaussian with zero-mean and unit-variance. Within each block, the transmission contains a training phase and a subsequent data transmission phase.
Here, we assume that a pilot sequence $\mathbf{s}_{j,k}$ of length $\tau$ is assigned to the $k$-th user of the $j$-th BS in the training phase.
And the remaining time resources, $T-\tau$, will be used for data transmission within a block.

Since \cite{wang2020extending} has already proven that the case with multiple antennas at the BSs differs from the single-antenna case by only a constant factor, we consider only the single-antenna case here for simplicity.
During the uplink training phase, each BS estimates the channels of users within its own cell based on the received pilot signals.
The received signal at the $j$-th BS during the training transmission can be represented by
\begin{equation}\label{equation_model1}
\mathbf{y}^{\mathrm{pilot}}_j=\sum_{k=0}^{K-1} h_{j,j,k}\mathbf{s}_{j,k} +\sum_{\substack{\Bar{j}=0\\ \Bar{j} \neq j}}^{J-1} \sqrt{\beta_{j,\Bar{j}}} \sum_{k=0}^{K-1} h_{j,\Bar{j},k}\mathbf{s}_{\Bar{j},k} +\mathbf{n}^{\mathrm{pilot}}_j,
\end{equation}
where $h_{j,\Bar{j},k} \sim \mathcal{CN}(0,1)$ denotes the channel coefficient from the $k$-th user in cell $\Bar{j}$ to BS $j$, and $\mathbf{y}^{\mathrm{pilot}}_j, \mathbf{n}^{\mathrm{pilot}}_j \in \mathbb{C}^{\tau\times 1}$ stand for the received signal and additive noise at BS $j$, respectively.
Moreover, the transmitted pilot sequence from user $k$ in cell $j$ is denoted by $\mathbf{s}_{j,k}\in \mathbb{C}^{\tau\times 1}$ with $\|\mathbf{s}_{j,k}\|^2=1$.
Then the collection of pilot sequences of all the users in cell $j$ can be denoted by $\mathbf{S}_j=[\mathbf{s}_{j,0},\ldots,\mathbf{s}_{j,K-1}]\in \mathbb{C}^{\tau\times K}, j=0,1,\ldots,J-1$, and the overall pilot matrix in the system becomes $\mathcal{S}=[\mathbf{S}_0,\ldots,\mathbf{S}_{J-1}]\in \mathbb{C}^{\tau\times JK}$ which contains $JK$ column vectors.

To simplify Eq. (\ref{equation_model1}) in form, we define $\beta_{j,j}=1, j=0,\ldots, J-1$.
This approach is reasonable as it indicates that the importance of intra-cell interference is the highest.
Then Eq. (\ref{equation_model1}) can be reduced to
\begin{equation}\label{equation_model2}
\mathbf{y}^{\mathrm{pilot}}_j=\sum_{\Bar{j}=0}^{J-1} \sqrt{\beta_{j,\Bar{j}}} \sum_{k=0}^{K-1} h_{j,\Bar{j},k}\mathbf{s}_{\Bar{j},k} +\mathbf{n}^{\mathrm{pilot}}_j.
\end{equation}

\subsection{Channel Estimation and Pilot Sequence Deign Problem}
At BS $j$, the least squares (LS) estimate of $h_{j,j,k}$ is given by
\begin{equation}
\hat{h}_{j,j,k}^{\mathrm{LS}}= \mathbf{s}_{j,k}^H \mathbf{y}^{\mathrm{pilot}}_j, j=0,\ldots,J-1,
\end{equation}
and the corresponding mean-squared-error (MSE) can be computed as \cite{hoydis2013massive, hassibi2003much, wang2020extending}
\begin{equation}\label{equation_MSEjk}
\mathrm{MSE}_{jk}^{\mathrm{LS}}= \mathbb{E}\left\{\|h_{j,j,k}-\hat{h}_{j,j,k}^{\mathrm{LS}}\|^2\right\} =\Psi_{j,k}+\Upsilon_{j,k}+ \sigma^2,
\end{equation}
where $\sigma^2$ is the noise variance, and for notational brevity we define
\begin{align}
\Psi_{j,k}=& \sum_{\substack{n=0\\ n \neq k}}^{K-1} \mathbf{s}_{j,k}^H \mathbf{s}_{j,n} \mathbf{s}_{j,n}^H \mathbf{s}_{j,k}, \\
\Upsilon_{j,k}=& \sum_{\substack{\Bar{j}=0\\ \Bar{j} \neq j}}^{J-1} \beta_{j,\Bar{j}} \sum_{n=0}^{K-1} \mathbf{s}_{j,k}^H \mathbf{s}_{\Bar{j},n} \mathbf{s}_{\Bar{j},n}^H \mathbf{s}_{j,k}, \label{equation_Psi}
\end{align}
which represent intra-cell and inter-cell interference in the training phase, respectively.

From Eqs. (\ref{equation_MSEjk}) - (\ref{equation_Psi}), the total intra-cell or inter-cell interference in the training phase of the system can be computed by
\begin{align}
I_\mathrm{intra}&= \sum_{j=0}^{J-1} \sum_{k=0}^{K-1} \Psi_{j,k} \notag\\
&= \sum_{j=0}^{J-1} \|\mathbf{S}_j^H\mathbf{S}_j\|_F^2 -JK,\\
I_\mathrm{inter}&= \sum_{j=0}^{J-1} \sum_{k=0}^{K-1} \Upsilon_{j,k} \notag\\
&= \sum_{m\neq n} \beta_{m, n} \|\mathbf{S}_m^H\mathbf{S}_n\|_F^2.
\end{align}

Next, by taking $\mathbf{B}$ into consideration, we define the extended Gram matrix as
\begin{equation}\label{equation_GB}
\mathbf{G_B}=
\left[
  \begin{array}{ccc}
    \sqrt{\beta_{0,0}}\mathbf{S}_0^H\mathbf{S}_0 &  \cdots & \sqrt{\beta_{0,J-1}}\mathbf{S}_0^H\mathbf{S}_{J-1} \\
    \vdots &   & \vdots \\
    \sqrt{\beta_{J-1,0}}\mathbf{S}_{J-1}^H\mathbf{S}_0 &  \cdots & \sqrt{\beta_{J-1,J-1}}\mathbf{S}_{J-1}^H\mathbf{S}_{J-1} \\
  \end{array}
\right],
\end{equation}
which is Hermitian and positive semi-definite.
Note that the entries of $\mathbf{G_B}$ are given by the weighted inner products of the corresponding sequences.
And the extended total squared correlation (ETSC) is defined as
\begin{equation}\label{equation_ETSC}
\mathrm{ETSC}(\mathcal{S},\mathbf{B})=\|\mathbf{G_B}\|^2_F,
\end{equation}
which characterizes the total interference in the system.

From (\ref{equation_MSEjk}), the sum MSE of all the users in the system in the training phase for LS estimate can be computed by
\begin{align} \label{equation_MSE}
\mathrm{MSE}^\mathrm{LS} =& \sum_{j=0}^{J-1} \sum_{k=0}^{K-1} \mathrm{MSE}_{jk}^{\mathrm{LS}} \notag \\
=& \sum_{j=0}^{J-1} \sum_{k=0}^{K-1} \Psi_{j,k}+\Upsilon_{j,k}+ \sigma^2 \notag \\
=& \|\mathbf{G_B}\|^2_F -JK +JK\sigma^2.
\end{align}

From Eq. (\ref{equation_MSE}), it can be seen that in order to minimize the overall interference in the system, we can only reduce the ETSC of the pilot sequence set, i.e., minimize pilot contamination, it is necessary and sufficient to minimize the ETSC.
Therefore, the pilot sequence design problem for multi-cell interference networks can be written as
\begin{align}\label{problem1}
    \mathcal{P}1: ~&\arg\min_{\mathcal{S}} \|\mathbf{G_B}\|^2_F, \\
    & \hbox{~~~~ s.t.} \|\mathbf{s}_{j,k}\|^2=1, \hbox{for all~} j,k. \notag
\end{align}

Peak-to-Average Power Ratio (PAPR) is an important metric for evaluating the transmitted signal in wireless communications. Typically, the higher the PAPR of a signal, the more severe the nonlinear distortion when passing through a power amplifier. The PAPR of the pilot sequence for the $k$-th user in the $j$-th cell is defined as \cite{litsyn2007peak}
\begin{equation}
\mathrm{PAPR}(\mathbf{s}_{j,k})=10 \log_{10} \frac{\max_t |s_{j,k}[t]|^2}{\frac{1}{\tau} \sum_{t=0}^{\tau-1} |s_{j,k}[t]|^2} (\mathrm{dB}),
\end{equation}
where $\mathbf{s}_{j,k}=[s_{j,k}[0],s_{j,k}[1],\ldots,s_{j,k}[\tau-1]]^T$.

A unimodular sequence refers to a sequence where the instantaneous power remains the same at all times, i.e.,
\begin{equation}\label{equation_unimodular}
|s_{j,k}[t]|^2=\frac{1}{\tau}, t=0,1,\ldots,\tau.
\end{equation}
This means that every element of the sequence has a constant magnitude, ensuring uniform power across all time instances.
Clearly, unimodular sequence have the lowest PAPR, meeting 0 dB.
When we require all users' pilot sequences to be unimodular sequences, the optimization problem $\mathcal{P}1$ will be modified as follows
\begin{align}\label{problem2}
    \mathcal{P}2: ~&\arg\min_{\mathcal{S}} \|\mathbf{G_B}\|^2_F, \\
    & \hbox{~~~~ s.t.} \left|{s}_{j,k}[t]\right|^2=\frac{1}{\tau}, \hbox{for all~} j,k,t. \notag
\end{align}

\subsection{Welch Bound and Extended Welch Bound}
In this subsection, we will briefly introduce the Welch bound and the previous extended Welch bound \cite{wang2020extending}.
In Section IV, we will demonstrate that the ETSC-MM algorithm proposed in this paper can meet these two theoretical bounds under certain specific parameters.

In the literature, the Welch bound characterizes the degree of orthogonality for a sequence set.
The Welch bound can be considered as the theoretical bound of TSC without taking into account the differences between intra-cell and inter-cell interference.
Let $\mathbf{B}$ be an all ones matrix, i.e., $\beta_{m,n}=1$.
Then the Gram matrix of $\mathcal{S}$ is
\begin{equation}
\mathbf{G}=\mathcal{S}^H\mathcal{S}.
\end{equation}
And $\|\mathbf{G}\|^2_F$ is called the total squared correlation (TSC) of $\mathcal{S}$ and is denoted $\mathrm{TSC}(\mathcal{S})=\|\mathbf{G}\|^2_F$.
The Welch bound \cite{welch1974lower} states that
\begin{equation}
\mathrm{TSC}(\mathcal{S}) \geq \frac{N^2}{\tau},
\end{equation}
where $N=JK$ represents the total number of users\footnote{The Welch bound has multiple forms. In this paper, we introduce only the form related to TSC.}.

To address the limitation of the Welch bound in not accounting for inter-cell interference, \cite{wang2020extending} proposed an extended Welch bound for two-cell.
Let $J=2,K\leq \tau \leq 2K$ and $\mathbf{B}=\left[
  \begin{array}{cc}
    1 & \beta \\
    \beta & 1 \\
  \end{array}
\right].$
Then extended Welch bound \cite{wang2020extending} states that
\begin{equation}
\mathrm{ETSC}(\mathcal{S}, \mathbf{B}) \geq \frac{2K^2(1+\beta)}{K+\beta(\tau-K)}.
\end{equation}

\section{A New Extended Welch Bound}
In this section, we will propose a new extended Welch bound and the necessary condition for the bound to meet.

The previous extended Welch bound in \cite{wang2020extending} provides a lower bound for the ETSC in the case of $J=2$ and $K\leq\tau$.
However, the theoretical bounds for arbitrary $J\in \mathbb{N}^+$ and $K>\tau$ have not been reported in the literature. In this subsection, we will present a new extended Welch bound to address this gap.
Before introducing the new bound we propose, we will first present a useful theorem, which is the cross form of the Inner Product Theorem \cite{welch1974lower, massey1991welch, massey1993welch}.
\begin{Theorem}[Cross-inner Product Theorem]\label{Theorem_CIPT}
Let $\mathbf{A,B}$ be two matrices with size $\tau\times K$, then we have
\begin{equation}
\sum_{m=0}^{K-1} \sum_{n=0}^{K-1} |\langle\mathbf{a}_m,\mathbf{b}_n\rangle|^2 = \sum_{\mu=0}^{\tau-1} \sum_{\nu=0}^{\tau-1} \langle\mathbf{a}^\mu,\mathbf{a}^\nu\rangle (\langle\mathbf{b}^\mu,\mathbf{b}^\nu\rangle)^*,
\end{equation}
where $\mathbf{a}_m,\mathbf{b}_m$ represent the $m$-th column vector of $\mathbf{A,B}$, and $\mathbf{a}^\mu,\mathbf{b}^\mu$ represent the $\mu$-th row vector of $\mathbf{A,B}$, respectively.
\end{Theorem}
\begin{IEEEproof}
We have
\begin{align*}
\sum_{m=0}^{K-1} \sum_{n=0}^{K-1} |\langle\mathbf{a}_m,\mathbf{b}_n\rangle|^2 = & \sum_{m=0}^{K-1} \sum_{n=0}^{K-1} \left|\sum_{\mu=0}^{\tau-1} a_m^\mu (b_n^\mu)^* \right|^2 \\
= & \sum_{m=0}^{K-1} \sum_{n=0}^{K-1} \sum_{\mu=0}^{\tau-1} \sum_{\nu=0}^{\tau-1} a_m^\mu (b_n^\mu)^* (a_m^\nu)^* b_n^\nu  \\
= & \sum_{\mu=0}^{\tau-1} \sum_{\nu=0}^{\tau-1} \sum_{m=0}^{K-1} \sum_{n=0}^{K-1} a_m^\mu (a_m^\nu)^* (b_n^\mu)^* b_n^\nu \\
= & \sum_{\mu=0}^{\tau-1} \sum_{\nu=0}^{\tau-1} \langle\mathbf{a}^\mu,\mathbf{a}^\nu\rangle (\langle\mathbf{b}^\mu,\mathbf{b}^\nu\rangle)^*.
\end{align*}
\end{IEEEproof}

\begin{Remark}
It is easy too see that if $\mathbf{A}=\mathbf{B}$, the Cross-inner product theorem reduces to the traditional Inner Product Theorem \cite{welch1974lower, massey1991welch, massey1993welch}, that is:
\begin{equation}
\sum_{m=0}^{K-1} \sum_{n=0}^{K-1} |\langle\mathbf{a}_m,\mathbf{a}_n\rangle|^2 = \sum_{\mu=0}^{\tau-1} \sum_{\nu=0}^{\tau-1} |\langle\mathbf{a}^\mu,\mathbf{a}^\nu\rangle|^2.
\end{equation}
\end{Remark}

By the Cross-inner Product Theorem, we can derive the extended Welch bound for the case where $K\geq \tau$.
\begin{Theorem}[The New Extended Welch Bound]\label{Theorem_NEWB}
Let $\mathcal{S}=[\mathbf{S}_0,\ldots,\mathbf{S}_{J-1}]$ be a sequence set for $J$ cells, where $\mathbf{S}_j=[\mathbf{s}_{j,0},\ldots, \mathbf{s}_{j,K-1}]$ and $\mathbf{s}_{j,k},k=0,\ldots, K-1$ are unimodular sequences.
Then we have
\begin{equation}\label{equation_EWB0}
\mathrm{ETSC}(\mathcal{S},\mathbf{B})\geq \frac{K^2}{\tau} \sum_{i=0}^{J-1} \sum_{j=0}^{J-1} \beta_{i,j},
\end{equation}
if $K\geq \tau$, and the inter-cell interference power factor matrix $\mathbf{B}$ is positive definite. And necessary condition for equality to hold in Eq.(\ref{equation_EWB0}) is that each $\mathbf{S}_j$ must be a WBE sequence set.
\end{Theorem}
\begin{IEEEproof}
By the Definition of ETSC, we have
\begin{equation*}
\mathrm{ETSC}(\mathcal{S},\mathbf{B})= \sum_{i=0}^{J-1} \sum_{j=0}^{J-1} \beta_{i,j} \sum_{m=0}^{K-1} \sum_{n=0}^{K-1} |\langle\mathbf{s}_{i,m},\mathbf{s}_{j,n}\rangle|^2.
\end{equation*}
As the \textit{Theorem} \ref{Theorem_CIPT}, we have
\begin{align}\label{equation_EWB}
\mathrm{ETSC}(\mathcal{S},\mathbf{B})\stackrel{(a)}{=}& \sum_{i=0}^{J-1} \sum_{j=0}^{J-1} \beta_{i,j} \sum_{\mu=0}^{\tau-1} \sum_{\nu=0}^{\tau-1} \langle\mathbf{s}_{i}^\mu,\mathbf{s}_{i}^\nu\rangle (\langle\mathbf{s}_{j}^\mu,\mathbf{s}_{j}^\nu\rangle)^* \notag\\
\stackrel{(b)}{=}& \sum_{\mu=0}^{\tau-1} \sum_{\nu=0}^{\tau-1} \sum_{i=0}^{J-1} \sum_{j=0}^{J-1} \beta_{i,j} \langle\mathbf{s}_{i}^\mu,\mathbf{s}_{i}^\nu\rangle (\langle\mathbf{s}_{j}^\mu,\mathbf{s}_{j}^\nu\rangle)^* \notag\\
\stackrel{(c)}{=}& \sum_{\mu=0}^{\tau-1} \sum_{i=0}^{J-1} \sum_{j=0}^{J-1} \beta_{i,j} \langle\mathbf{s}_{i}^\mu,\mathbf{s}_{i}^\mu\rangle (\langle\mathbf{s}_{j}^\mu,\mathbf{s}_{j}^\mu\rangle)^* \notag\\
&+ \sum_{\mu\neq\nu} \sum_{i=0}^{J-1} \sum_{j=0}^{J-1} \beta_{i,j} \langle\mathbf{s}_{i}^\mu,\mathbf{s}_{i}^\nu\rangle (\langle\mathbf{s}_{j}^\mu,\mathbf{s}_{j}^\nu\rangle)^*,
\end{align}
where $(a)$ follows from \textit{Theorem} \ref{Theorem_CIPT}.
By interchanging the order of summation, we obtain $(b)$.
By separating the terms where $\mu=\nu$, we obtain $(c)$.

Define $f_{\mu,\nu}(\mathbf{B})=\sum_{i=0}^{J-1} \sum_{j=0}^{J-1} \beta_{i,j} \langle\mathbf{s}_{i}^\mu,\mathbf{s}_{i}^\nu\rangle (\langle\mathbf{s}_{j}^\mu,\mathbf{s}_{j}^\nu\rangle)^*$.
Then we have
\begin{equation}\label{equation_fB}
    f_{\mu,\nu}(\mathbf{B})=\mathbf{x}^H\mathbf{B}\mathbf{x},
\end{equation}
where $\mathbf{x}=[\langle\mathbf{s}_{0}^\mu,\mathbf{s}_{0}^\nu\rangle, \ldots,\langle\mathbf{s}_{J-1}^\mu,\mathbf{s}_{J-1}^\nu\rangle]^H$.
Since $\mathbf{B}$ is a positive definite matrix, $f_{\mu,\nu}(\mathbf{B})\geq 0$, with equality attained when $\mathbf{x}=\mathbf{0}_J$.
Therefore, we have
\begin{align}\label{equation_EWB2}
\mathrm{ETSC}(\mathcal{S},\mathbf{B})
\stackrel{(d)}{=}& \sum_{\mu=0}^{\tau-1} \sum_{i=0}^{J-1} \sum_{j=0}^{J-1} \beta_{i,j} \langle\mathbf{s}_{i}^\mu,\mathbf{s}_{i}^\mu\rangle (\langle\mathbf{s}_{j}^\mu,\mathbf{s}_{j}^\mu\rangle)^*  \notag\\
&+ \sum_{\mu\neq\nu} f_{\mu,\nu}(\mathbf{B}) \notag\\
\stackrel{(e)}{\geq}& \frac{K^2}{\tau} \sum_{i=0}^{J-1} \sum_{j=0}^{J-1} \beta_{i,j}.
\end{align}
Substituting Eq. (\ref{equation_fB}) into Eq. (\ref{equation_EWB}), we obtain $(d)$.
From the property of positive definite matrix $\mathbf{B}$, we obtain $(e)$.
Note that the above inequality is met with equality when $\mathbf{s}_j^{\mu}$ and $\mathbf{s}_j^{\nu}$ are orthogonal for all $j=0,\ldots,J-1,\mu\neq \nu$.
This leads to the necessary condition for equality, which is that each $\mathbf{S}_j$ must be a WBE sequence set.
\end{IEEEproof}

\begin{Remark}
The previous extended Welch bound focuses on the case where $K\leq\tau$ and $J=2$.
In contrast, our proposed new extended Welch bound addresses the scenario where $K\geq\tau$ and $J\in \mathbb{N}^+$.
They provide lower bounds for ETSC in different circumstances.
Besides, the lower bound for ETSC in the case where $K\leq \tau$ and $J\geq3$ remains an open question.
\end{Remark}

\begin{Remark}
When $\mathbf{B}$ is a matrix of all ones, the new extended Welch bound reduces to the traditional Welch bound \cite{welch1974lower} in the case of unimodular sequences, which is given by
\begin{equation}
\mathrm{ETSC}(\mathcal{S},\mathbf{B}) \geq \frac{N^2}{\tau},
\end{equation}
where $N=JK$ and $\mathbf{B}=\mathbf{1}_{J\times J}$.
\end{Remark}

\section{ETSC Minimization via Majorization-Minimization}
In the previous section, we discussed the theoretical lower bound and the necessary condition for the bound to meet in the case where $\tau\leq K, J\in \mathbb{N}^+$, and $\mathbf{B}$ is a positive definite matrix.
Combining Wang's work \cite{wang2020extending}, we find that there remains a lack of methods for generating sequences in the following two cases: a) $\tau\geq K,J\geq 3$; b) $\tau \leq K, J\in \mathbb{N}^+$ and $\mathbf{B}$ is not a positive definite matrix.
Therefore, the development of a general sequence design method is encouraged, which should at least address the sequence generation problem for the two cases mentioned above.
In this section, we first introduce the general MM framework briefly and then develop a simple algorithm for the ETSC minimization problem based on the general MM framework.

\subsection{The MM Framework}
The MM method can be applied to solve the following general optimization problem:
\begin{align}\label{problem_general}
    \mathcal{P}3: ~&\arg\min_{\mathbf{x}} f(\mathbf{x}), \\
    & \hbox{~~~~ s.t.} \mathbf{x} \in \mathcal{X}, \notag
\end{align}
where $f$ is differentiable on the whole $\mathbb{C}$ space and $\mathcal{X}$ is some constraint set.
Rather than minimizing $f(\mathbf{x})$ directly, we consider successively solving a series of simple optimization problems.
The algorithm initializes at some feasible starting point $\mathbf{x}^{(0)}$, and then iterates as $\mathbf{x}^{(1)}, \mathbf{x}^{(2)}, \ldots$ until some convergence criterion is met.
For any iteration, say, the $l$-th iteration, the update rule is
\begin{equation}
\mathbf{x}^{(l+1)}=\arg\min_{\mathbf{x}\in \mathcal{X}} u(\mathbf{x}, \mathbf{x}^{(l)}),
\end{equation}
where $\mathbf{x}^{(l)})$ is the point generated by the algorithm at iteration $l$, and $u(\mathbf{x}, \mathbf{x}^{(l)})$ is the majorization function of $f(\mathbf{x})$ at $\mathbf{x}^{(l)}$.
Formally, the function $u(\mathbf{x}, \mathbf{x}^{(l)})$ is called to majorize the function $f(\mathbf{x})$ at $\mathbf{x}^{(l)}$ provided
\begin{enumerate}
    \item $u(\mathbf{x}, \mathbf{x}^{(l)}) \geq f(\mathbf{x}), \forall \mathbf{x} \in \mathcal{X}$;
    \item $u(\mathbf{x}^{(l)}, \mathbf{x}^{(l)}) = f(\mathbf{x}^{(l)}), \forall \mathbf{x}$.
\end{enumerate}
In other words, function $u(\mathbf{x}, \mathbf{x}^{(l)})$ is an upper bound of $f(\mathbf{x})$ over $\mathcal{X}$ and coincides with $f(\mathbf{x})$ at $\mathbf{x}^{(l)}$.

To summarize, to minimize $f(\mathbf{x})$ over $\mathcal{X}$, the main steps of the majorization-minimization framework are
\begin{enumerate}
    \item Find a feasible point $\mathbf{x}^{(0)}$ and set $l=0$.
    \item Construct a majorization function $u(\mathbf{x}, \mathbf{x}^{(l)})$ of $f(\mathbf{x})$ at $\mathbf{x}^{(l)}$.
    \item Let $\mathbf{x}^{(l+1)}=\arg\min_{\mathbf{x}\in \mathcal{X}} u(\mathbf{x}, \mathbf{x}^{(l)})$.
    \item If some convergence criterion is met, exit; otherwise, set $l=l+1$ and go to step (2).
\end{enumerate}
It is easy to show that with this framework, the objective value is
decreased monotonically at every iteration, i.e.,
\begin{equation}
f({\bf x}^{(l+1)})\!\leq\! u({\bf x}^{(l+1)}, {\bf x}^{(l)})\!\leq\! u({\bf x}^{(l)}, {\bf x}^{(l)})\!=\!f({\bf x}^{(l)}).
\end{equation}
The monotonicity makes MM framework very stable in practice.

\subsection{ETSC-MM Algorithm}
To solve the problem $\mathcal{P}1$ via majorization-minimization, the key step is to find a majorization function of the objective such that the majorized problem is easy to solve.
There is a useful lemma which can be employed to construct the majorization function \cite{song2015optimization}.
\begin{Lemma}[\cite{song2015optimization}] \label{lemma1}
Let $\mathbf{L}$ be an $n\times n$ Hermitian matrix and $\mathbf{M}$ be another $n\times n$ Hermitian matrix such that $\mathbf{M}
\succeq \mathbf{L}$.
Then for any point $\mathbf{x}_0 \in \mathbb{C}^n$, $\mathbf{x}^H\mathbf{Mx} +\mathrm{Re}(\mathbf{x}^H(\mathbf{L-M})\mathbf{x}_0) +\mathbf{x}_0^H(\mathbf{M-L})\mathbf{x}_0$ is a majorization function of the quadratic function $\mathbf{x}^H\mathbf{Lx}$ at $\mathbf{x}_0$.
\end{Lemma}

The objective of the problem $\mathcal{P}1$ is quartic with respect to $\mathcal{S}$. To find a majorization function, some reformulations are necessary.
For convenience, we will combine the cell index and user index, and rewrite equation (\ref{equation_GB}) as follow:
\begin{equation}\label{equation_GB2}
\mathbf{G_B}=
\left[
  \begin{array}{ccc}
    \sqrt{w_{0,0}}\mathbf{x}_0^H\mathbf{x}_0 & \cdots & \sqrt{w_{0,N-1}}\mathbf{x}_0^H\mathbf{x}_{N-1} \\
    \vdots &    & \vdots \\
    \sqrt{w_{N-1,0}}\mathbf{x}_{N-1}^H\mathbf{x}_0 & \cdots & \sqrt{w_{N-1,N-1}}\mathbf{x}_{N-1}^H\mathbf{x}_{N-1} \\
  \end{array}
\right],
\end{equation}
where $N=JK$, $\mathbf{x}_n=\mathbf{s}_{j,k}$, $n=jK+k,j=0,\ldots,J-1, k=0,\ldots,K-1$, and $w_{m,n}=\beta_{\lfloor\frac{m}{K}\rfloor, \lfloor\frac{n}{K}\rfloor}$.

Let us first stack the sequences $\mathbf{x}_n, n=0,\ldots,N-1$ together and denote it by $\mathbf{x}$, i.e.,
\begin{equation}
\mathbf{x}=[\mathbf{x}^T_0,\mathbf{x}^T_1,\ldots, \mathbf{x}^T_{N-1}]^T,
\end{equation}
then we have
\begin{equation}
\mathbf{x}_n=\mathbf{T}_n\mathbf{x}, n=0,\ldots,N-1,
\end{equation}
where $\mathbf{T}_n$ is an $\tau\times \tau N$ block selection matrix defined as
\begin{equation}
\mathbf{T}_n=[\mathbf{0}_{\tau\times (n-1)\tau}, \mathbf{I}_\tau, \mathbf{0}_{\tau\times (N-n)\tau}].
\end{equation}
We then note that $[\mathbf{G_B}]_{ij}$ can be written as
\begin{equation}
[\mathbf{G_B}]_{ij}=\sqrt{w_{ij}}\mathbf{x}^H\mathbf{T}_i^H\mathbf{T}_j\mathbf{x}.
\end{equation}
Let $\mathbf{U}_{ij}=\mathbf{T}_i^H\mathbf{T}_j$, we have
\begin{align}
\|\mathbf{G_B}\|^2_F=& \sum_{i=1}^N\sum_{j=1}^N w_{ij} |\mathbf{x}^H \mathbf{U}_{ij} \mathbf{x}|^2 \notag\\
=& \sum_{i=1}^N\sum_{j=1}^N w_{ij} |\mathrm{Tr}(\mathbf{U}_{ij} \mathbf{X})|^2 \notag\\
=& \sum_{i=1}^N\sum_{j=1}^N w_{ij} \mathrm{vec}(\mathbf{X})^H \mathrm{vec}(\mathbf{U}_{ij}) \mathrm{vec}(\mathbf{U}_{ij})^H \mathrm{vec}(\mathbf{X}), \label{equation_GB_norm}
\end{align}
where $\mathrm{vec}(\mathbf{X})$ is a column vector consisting of all the columns of $\mathbf{X}$ stacked, and $\mathbf{X}=\mathbf{x}\mathbf{x}^H$.

By using (\ref{equation_GB_norm}), problem $\mathcal{P}1$ can be rewritten as
\begin{align}\label{problem4}
    \mathcal{P}4: ~&\arg\min_{\mathcal{S}} \mathrm{vec}(\mathbf{X})^H \mathbf{\Lambda}_1 \mathrm{vec}(\mathbf{X}), \\
    & \hbox{~~~~ s.t.} \|\mathbf{x}_{n}\|^2=1, n=0,\ldots,N-1, \notag
\end{align}
where
\begin{equation}\label{equation_Lambda1}
\mathbf{\Lambda}_1=\sum_{i=1}^N\sum_{j=1}^N w_{ij}  \mathrm{vec}(\mathbf{U}_{ij}) \mathrm{vec}(\mathbf{U}_{ij})^H.
\end{equation}

Then given $\mathbf{X}^{(l)}$ at iteration $l$, by using \textit{Lemma} \ref{lemma1}, we can majorize the objective of (\ref{problem1}) by a quadratic function and the majorized problem after the first majorization step is given by the following function at $\mathbf{X}^{(l)}$
\begin{align}\label{equation_u1}
u_1(\mathbf{X}, \mathbf{X}^{(l)})=&\mathrm{vec}(\mathbf{X})^H (\mathbf{\Lambda}_1\mathbf{I}) \mathrm{vec}(\mathbf{X}) \notag\\
&+ 2 \operatorname{Re}\left(\mathrm{vec}(\mathbf{X})^H(\mathbf{\Lambda}_1-\lambda_1\mathbf{I}) \mathrm{vec}(\mathbf{X}^{(l)})\right) \notag\\
&+(\mathrm{vec}(\mathbf{X}^{(l)}))^H(\lambda_1\mathbf{I}-\mathbf{\Lambda}_1) \mathrm{vec}(\mathbf{X}^{(l)}),
\end{align}
where $\lambda_{1}$ is the largest eigenvalue of $\mathbf{\Lambda}_1$.

Since the energies of $\mathbf{x}_n,n=0,\ldots,N-1$ are equal to 1, it is easy to see that the first term of (\ref{equation_u1}) is a constant.
After ignoring the constant terms, the majorized problem of $\mathcal{P}4$ is given by
\begin{align}\label{problem5}
    \mathcal{P}5: ~&\arg\min_{\mathcal{S}} \mathrm{Re}\left(\mathrm{vec}(\mathbf{X})^H(\mathbf{\Lambda}_1-\lambda_1\mathbf{I}) \mathrm{vec}(\mathbf{X}^{(l)})\right), \\
    & \hbox{~~~~ s.t.} \|\mathbf{x}_{n}\|^2=1, n=0,\ldots,N-1. \notag
\end{align}

By substituting $\mathbf{\Lambda}_1$ in (\ref{equation_Lambda1}) into (\ref{problem5}), we can derive the intrinsic relationship between problem $\mathcal{P}5$ and $\mathbf{x}$, given in (\ref{equation_xLambda2x}), where
\begin{equation}\label{equation_Lambda2}
\mathbf{\Lambda}_2=\sum_{i=1}^N\sum_{j=1}^N w_{ij} \mathrm{vec}(\mathbf{U}_{ij})^H \mathrm{vec}(\mathbf{X}^{(l)}) \mathbf{U}_{ij}.
\end{equation}
\begin{figure*}[ht]
\begin{align}\label{equation_xLambda2x}
\mathrm{Re}\left(\mathrm{vec}(\mathbf{X})^H(\mathbf{\Lambda}_1-\lambda_1\mathbf{I}) \mathrm{vec}(\mathbf{X}^{(l)})\right) =& \mathrm{Re} \left(\mathrm{vec}(\mathbf{X})^H\left(\sum_{i=1}^N\sum_{j=1}^N w_{ij}  \mathrm{vec}(\mathbf{U}_{ij}) \mathrm{vec}(\mathbf{U}_{ij})^H-\lambda_1\mathbf{I}\right) \mathrm{vec}(\mathbf{X}^{(l)})\right) \notag\\
=& \mathrm{Re}\left( \sum_{i=1}^N\sum_{j=1}^N w_{ij} \mathrm{vec}(\mathbf{X})^H \mathrm{vec}(\mathbf{U}_{ij}) \mathrm{vec}(\mathbf{U}_{ij})^H \mathrm{vec}(\mathbf{X}^{(l)})  -\lambda_1 \mathrm{vec}(\mathbf{X})^H \mathrm{vec}(\mathbf{X}^{(l)}) \right) \notag\\
=& \mathrm{Re}\left( \sum_{i=1}^N\sum_{j=1}^N w_{ij} \mathbf{x}^H \mathbf{U}_{ij} \mathbf{x} \cdot \mathrm{vec}(\mathbf{U}_{ij})^H \mathrm{vec}(\mathbf{X}^{(l)}) -\lambda_1 \mathbf{x}^H\mathbf{x}^{(l)}(\mathbf{x}^{(l)})^H\mathbf{x} \right) \notag\\
=& \mathrm{Re}\left( \mathbf{x}^H\left(\sum_{i=1}^N\sum_{j=1}^N w_{ij} \mathrm{vec}(\mathbf{U}_{ij})^H \mathrm{vec}(\mathbf{X}^{(l)}) \mathbf{U}_{ij} - \lambda_1 \mathbf{x}^{(l)}(\mathbf{x}^{(l)})^H \right) \mathbf{x} \right) \notag\\
=& \mathrm{Re}\left( \mathbf{x}^H(\mathbf{\Lambda}_2 - \lambda_1 \mathbf{x}^{(l)}(\mathbf{x}^{(l)})^H)\mathbf{x} \right),
\end{align}
\end{figure*}

It should be noted that $\mathbf{x}^H(\mathrm{vec}(\mathbf{U}_{ij})^H \mathrm{vec}(\mathbf{X}^{(l)}) \mathbf{U}_{ij})\mathbf{x}$ and $\mathbf{x}^H(\mathrm{vec}(\mathbf{U}_{ji})^H \mathrm{vec}(\mathbf{X}^{(l)}) \mathbf{U}_{ji})\mathbf{x}$ are a pair of complex conjugates.
Hence, $\mathbf{x}^H\mathbf{\Lambda}_2\mathbf{x}$ is always a real number, and problem $\mathcal{P}5$ can be rewritten as
\begin{align}\label{problem6}
    \mathcal{P}6: ~&\arg\min_{\mathcal{S}} \mathbf{x}^H(\mathbf{\Lambda}_2 - \lambda_1 \mathbf{x}^{(l)}(\mathbf{x}^{(l)})^H)\mathbf{x}, \\
    & \hbox{~~~~ s.t.} \|\mathbf{x}_{n}\|^2=1, n=0,\ldots,N-1, \notag
\end{align}
where $\mathbf{\Lambda}_2$ is given in (\ref{equation_Lambda2}).

Since the majorized problem $\mathcal{P}6$ is still hard to solve directly, we propose to majorize the objective function at $\mathbf{x}^{(l)}$ again to further simplify the problem that we need to solve at each iteration.
Similarly, to construct a majorization function of the quadratic objective in (\ref{problem6}), we need to find a matrix $\mathbf{M}$ such that $\mathbf{M}\succeq \mathbf{\Lambda}_2 - \lambda_1 \mathbf{x}^{(l)}(\mathbf{x}^{(l)})^H$ and a straightforward choice may be $\mathbf{M}=\lambda_{\max}(\mathbf{\Lambda}_2 - \lambda_1 \mathbf{x}^{(l)}(\mathbf{x}^{(l)})^H) \mathbf{I}$.
But to compute the maximum eigenvalue, it will be computationally expensive. To maintain the computational efficiency of the algorithm, we propose to use some upper bound of $\lambda_{\max}(\mathbf{\Lambda}_2 - \lambda_1 \mathbf{x}^{(l)}(\mathbf{x}^{(l)})^H)$ that can be easily computed.
Let $\lambda_2$ be the maximum eigenvalue of $\mathbf{\Lambda}_2$.
Clearly, $\lambda_2$ is also an upper bound for $\lambda_{\max}(\mathbf{\Lambda}_2 - \lambda_1 \mathbf{x}^{(l)}(\mathbf{x}^{(l)})^H)$.

Now, by choosing $\mathbf{M}=\lambda_2\mathbf{I}$ in \textit{Lemma} \ref{lemma1}, the objective in (\ref{problem6}) is majorized by
\begin{align}\label{equation_u2}
u_2(\mathbf{x}, \mathbf{x}^{(l)})=&\mathbf{x}^H (\lambda_2\mathbf{I}) \mathbf{x} \notag\\
&+ 2 \operatorname{Re}\left(\mathbf{x}^H(\mathbf{\Lambda}_2 -\lambda_1 \mathbf{x}^{(l)}(\mathbf{x}^{(l)})^H -\lambda_2\mathbf{I}) \mathbf{x}^{(l)}\right) \notag\\
&+(\mathbf{x}^{(l)})^H(\lambda_2\mathbf{I}-\mathbf{\Lambda}_2 +\lambda_1 \mathbf{x}^{(l)}(\mathbf{x}^{(l)})^H) \mathbf{x}^{(l)}
\end{align}

By ignoring identities and constant terms, optimization problem (\ref{problem1}) can be transformed into the following optimization problem.
\begin{align}\label{problem7}
    \mathcal{P}7: ~&\arg\min_{\mathcal{S}} \operatorname{Re}\left(\mathbf{x}^H(\mathbf{\Lambda}_2 -\lambda_1 \mathbf{x}^{(l)}(\mathbf{x}^{(l)})^H -\lambda_2\mathbf{I}) \mathbf{x}^{(l)}\right), \\
    & \hbox{~~~~ s.t.} \|\mathbf{x}_{n}\|^2=1, n=0,\ldots,N-1, \notag
\end{align}
where $\mathbf{\Lambda}_2$ is given in (\ref{equation_Lambda2}) and $\lambda_2$ is the maximum eigenvalue of $\mathbf{\Lambda}_2$.

Note that optimization problem $\mathcal{P}7$ can be rewritten as
\begin{align}\label{problem8}
    \mathcal{P}8: ~&\arg\min_{\mathcal{S}} \left\|\mathbf{x}- \mathbf{y}\right\|^2, \\
    & \hbox{~~~~ s.t.~} \mathbf{x}=[\mathbf{x}^T_0,\mathbf{x}^T_1,\ldots, \mathbf{x}^T_N,]^T \hbox{~and~} \|\mathbf{x}_n\|^2=1, \notag
\end{align}
where
\begin{equation}\label{equation_y}
\mathbf{y}=-(\mathbf{\Lambda}_2 -\lambda_1 \mathbf{x}^{(l)}(\mathbf{x}^{(l)})^H -\lambda_2\mathbf{I}) \mathbf{x}^{(l)}.
\end{equation}

It is clear that problem $\mathcal{P}8$ is separable in the sub-sequences of $\mathbf{x}$ and the solution of the problem is given by
\begin{equation}\label{equation_solution1}
\mathbf{x}_n=\frac{\mathbf{y}_n}{\|\mathbf{y}_n\|}, n=0,\ldots,N-1,
\end{equation}
where $\mathbf{y}_n=[y[n\tau],y[n\tau+1],\ldots,y[(n+1)\tau-1]]^T$.

So far, we have presented the complete solution process for $\mathcal{P}1$. The process for solving $\mathcal{P}2$ is similar, with the only difference being the additional constraint given by Eq. (\ref{equation_unimodular}). Therefore, we only need to modify Eq. (\ref{equation_solution1}) to the following equation
\begin{equation}\label{equation_solution2}
{x}_n[t]=e^{i \varphi(y[n\tau+t])}, n=0,\ldots,N-1, t=0,\ldots,\tau-1.
\end{equation}

According to the general steps of the majorization minimization framework, we can now implement the algorithm in a straightforward way, that is at each iteration, we compute $\mathbf{y}$ according to (\ref{equation_y}) and update $\mathbf{x}$ via (\ref{equation_solution1}) or (\ref{equation_solution2}).
The overall algorithm is summarized in \textbf{Algorithm 1}\footnote{The MATLAB codes used in numerical experiments of this paper can be downloaded from the author's GitHub website: \\https://github.com/ZhiGuMath/NonOrthogonalSequence.}.

\begin{table}[ht]
\centering
\label{Algorithm_ETSC}
\begin{tabular}{p{8cm}}
\hline\hline
\textbf{Algorithm 1:} The ETSC-MM Algorithm  \\ \hline
\textbf{Input:} Sequence length $\tau$; Number of cells $J$; Number of users in single cell $K$; Inter-cell interference power factor matrix $\mathbf{B}$; Maximum iterations number $\mathrm{MaxItrNum}$;
Initial point $\mathbf{x}^{(0)}$ is a random sequence. Initialize $l$ to $0$. \\
\textbf{Step 1:} Calculate $\mathbf{\Lambda_1}$ by Eq. (\ref{equation_Lambda1}) and its the largest eigenvalue $\lambda_1$. \\
\textbf{Step 2:} Calculate $\mathbf{\Lambda_2}$ by Eq. (\ref{equation_Lambda2}) and its the largest eigenvalue $\lambda_2$. \\
\textbf{Step 3:} Calculate $\mathbf{y}$ by Eq. (\ref{equation_y}).\\
\textbf{Step 4:} Update $\mathbf{x}^{(l+1)}$ by Eq. (\ref{equation_solution1}) or Eq. (\ref{equation_solution2}). \\
\textbf{Iteration:} Repeat Steps 2-4 until some stop criterion is satisfied, e.g., $\|\mathbf{x}^{(l)}-\mathbf{x}^{(l-1)}\|^2\leq \varepsilon$, where $\varepsilon$ is a predefined threshold, or $l>\mathrm{MaxItrNum}$. Otherwise, update $l=l+1$ and continue the iterations. \\
\textbf{Output:} Set the optimized sequence as $\mathbf{a}$. \\
\hline\hline
\end{tabular}
\end{table}

\subsection{Computational Complexity of ETSC-MM}
In this section, we will provide a computational complexity analysis of the ETSC-MM algorithm and present some techniques for reducing its complexity.
From \textbf{Algorithm 1}, it is evident that the computational cost of the ETSC-MM algorithm primarily arises from three components: the computation of the eigenvalue $\lambda_1$, the computation of the eigenvalue $\lambda_2$, and other matrix multiplications.

The following theorem provides an analytical expression for the maximum eigenvalue of matrix $\Lambda_1$, which reduces the computational cost related to $\lambda_1$ in \textbf{Algorithm 1} to $\mathcal{O}(1)$.

\begin{Theorem}
For any given weight coefficients $w_{m,n}$, the maximum eigenvalue of matrix $\Lambda_1$ is always $\tau$.
\end{Theorem}
\begin{IEEEproof}
Clearly, the eigenvalues of matrix $\mathrm{vec}(\mathbf{U}_{ij}) \mathrm{vec}(\mathbf{U}_{ij})^H$ are $\mathrm{vec}(\mathbf{U}_{ij})^H \mathrm{vec}(\mathbf{U}_{ij})=\tau$ and 0, for all $i,j$.
Therefore, the eigenvalues of matrix $\mathbf{\Lambda}_1$ are $\tau w_{i,j}$ and 0.
Note that $w_{i,i}=1$, so the maximum eigenvalue of $\mathbf{\Lambda}_1$ is $\tau$.
\end{IEEEproof}

To efficiently compute the eigenvalues of $\mathbf{\Lambda}_2$, we need the following lemma.
\begin{Lemma}[\cite{horn1994topics}]
Let $\mathbf{A}$ and $\mathbf{B}$ be square matrices of size $M$ and $N$, respectively. Let $\lambda_1,\ldots,\lambda_M$ be the eigenvalues of $\mathbf{A}$ and $\mu_1,\ldots,\mu_N$ be those of $\mathbb{B}$. Then the eigenvalues of $\mathbf{A}\otimes \mathbf{B}$ are $\lambda_i\mu_j, i=1,\ldots,M,j=1,\ldots,N$.
\end{Lemma}

By simplifying Eq. (\ref{equation_Lambda2}), we have
\begin{align}\label{equation_Lambda2_2}
\mathbf{\Lambda}_2=& \sum_{i=1}^N\sum_{j=1}^N w_{ij} \mathrm{vec}(\mathbf{U}_{ij})^H \mathrm{vec}(\mathbf{X}^{(l)}) \mathbf{U}_{ij} \notag \\
=& \sum_{i=1}^N\sum_{j=1}^N \left(w_{ij} (\mathbf{x}^{(l)})^H \mathbf{U}_{ij} \mathbf{x}^{(l)}\right)^* \mathbf{U}_{ij} \notag\\
=& \mathbf{W} \odot ((\mathcal{S}^{(l)})^H\mathcal{S}^{(l)})^* \otimes \mathbf{I}_\tau,
\end{align}
where $\mathbf{W}=\mathbf{B}\otimes \mathbf{1}_K$.
Through Eq. (\ref{equation_Lambda2_2}) and Lemma 2, the maximum eigenvalue of matrix $\mathbf{\Lambda}_2$ is the same as that of matrix $\mathbf{W} \odot ((\mathcal{S}^{(l)})^H\mathcal{S}^{(l)})$. Since matrix $\mathbf{W} \odot ((\mathcal{S}^{(l)})^H\mathcal{S}^{(l)})$ is of size $N \times N$, the complexity of calculating its maximum eigenvalue is $\mathcal{O}(N^3)$.

It is evident from \textbf{Algorithm 1} that the highest computational complexity for matrix multiplication is $\mathcal{O}((\tau N)^2)$, which occurs during the calculation of Eq. (\ref{equation_y}).
This computational complexity can be even higher than that for calculating the eigenvalues of $\mathbf{\Lambda}_2$. Fortunately, we can utilize the sparseness of matrix $\mathbf{\Lambda}_2$ to reduce the computational cost of Eq. (\ref{equation_y}).
Firstly, Eq. (\ref{equation_y}) can be rewritten as
\begin{align}
\mathbf{y}=& -(\mathbf{\Lambda}_2 -\lambda_1 \mathbf{x}^{(l)}(\mathbf{x}^{(l)})^H -\lambda_2\mathbf{I}) \mathbf{x}^{(l)} \notag\\
=& -\mathbf{\Lambda}_2\mathbf{x}^{(l)} +\lambda_1 \mathbf{x}^{(l)}(\mathbf{x}^{(l)})^H\mathbf{x}^{(l)} +\lambda_2 \mathbf{x}^{(l)} \notag\\
=& -\mathbf{\Lambda}_2\mathbf{x}^{(l)} + (N\lambda_1 +\lambda_2) \mathbf{x}^{(l)}
\end{align}
Clearly, the computational complexity of the above equation is the number of non-zero elements in matrix $\mathbf{\Lambda}_2$, which is $\mathcal{O}(N^2 \tau)$.

\subsection{Convergence Analysis}
The ETSC-MM algorithm given in \textbf{Algorithm 1} is based on the general majorization-minimization framework, thus according to Section III-A, we know that the sequence of objective values (i.e., ETSC of sequence set) evaluated at $\mathbf{x}^{(l)}$ generated by the algorithm is nonincreasing.
And it is easy to see that the objective value is bounded below by 0, thus the sequence of objective values is guaranteed to converge to a finite value.

\subsection{Acceleration Scheme}
The squared iterative method (SQUAREM) is the method to accelerate any Expectation-Maximization (EM) algorithm \cite{varadhan2008simple}. It seeks to approximate Newton's method for finding a fixed point of the EM algorithm map and generally achieves super-linear convergence. Since SQUAREM only requires the EM updating map, it can be readily applied to any EM-type algorithms.
Since MM is a generalization of EM and the update rule is just a fixed-point iteration, SQUAREM can be readily applied to MM algorithms.
There are some successful applications of SQUAREM to accelerate MM, as demonstrated in cases such as \cite{song2015optimization, song2015sequence, song2016sequence, zhao2016unified}, within the field of sequence design.
Due to space limitations, we will not detail the application of SQUAREM to accelerate the ETSC-MM algorithm. Interested readers can refer to \cite{song2015optimization, song2015sequence, song2016sequence, zhao2016unified} for insights into using SQUAREM with the ETSC-MM algorithm.

\section{Numerical Experiments}
This section evaluates the performance of the ETSC-MM algorithm and the channel estimation performance of the optimal sequence set.
All experiments were performed on a PC with a 2.10GHz i7–12700 CPU and 16GB RAM.

\subsection{Performance Analysis of Theoretical Construction}
\begin{figure}[ht]
  \centering
  \includegraphics[width=8cm]{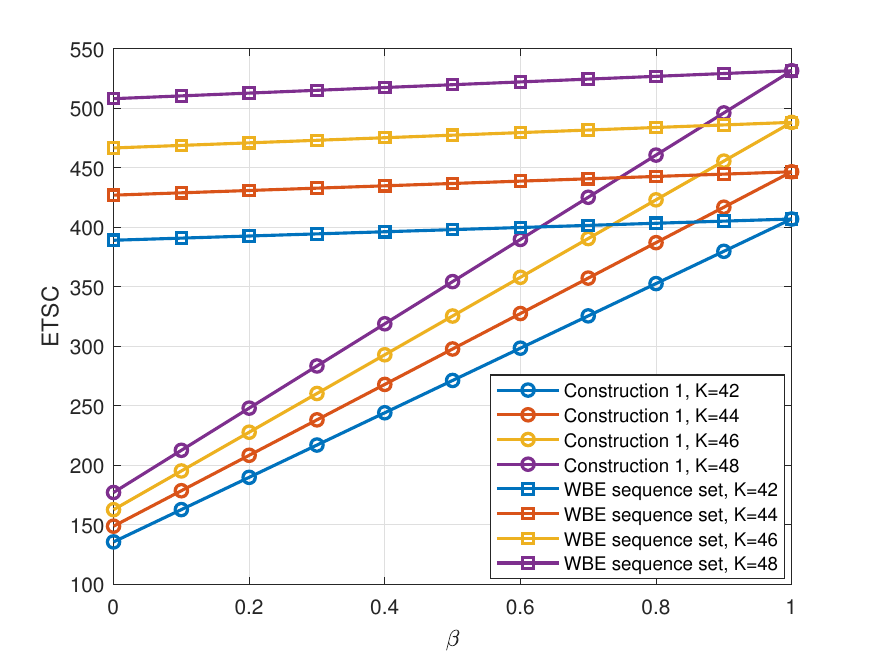}
  \caption{ETSC of sequence sets generated by \textit{Theorem} \ref{Theorem_NEWB} and the WBE sequence set with $\tau=39$, $K=42,44,46,48$, $J=3$ and different $\mathbf{B}$.}\label{fig_bound_duibi}
\end{figure}

In this subsection, we present the sequence sets generated by \textit{Theorem} \ref{Theorem_NEWB} under various parameters and compare them with the WBE sequence sets.
Here, we set the sequence length $\tau=39$, the number of users $K=42,44,46,48$, the number of cells $J=3$, and the interference power factor matrix $\mathbf{B}$ defined as follows
$$
\mathbf{B}=
\left[
  \begin{array}{ccc}
    1 & \beta & \beta \\
    \beta & 1 & \beta \\
    \beta & \beta & 1 \\
  \end{array}
\right], \beta=0,0.1,\ldots,1.
$$
Fig. \ref{fig_bound_duibi} compares the ETSC of the sequence sets generated by \textit{Theorem} \ref{Theorem_NEWB} and the WBE sequence sets with different interference power factor matrices $\mathbf{B}$.
It shows that when $\beta=1$, the ETSC of the sequence set generated by \textit{Theorem} \ref{Theorem_NEWB} is equal to that of the WBE sequence set, as the interference strength within the cell and between cells is the same in such case. Furthermore, in other instances, the sequences generated by \textit{Theorem} \ref{Theorem_NEWB} exhibit lower ETSC.

In the engineering context considered in this paper, there are two common types of positive definite matrices. Proofs of their positive definiteness can be found in \cite{golub2013matrix}. For brevity, the detailed proofs are omitted here.
\begin{enumerate}
    \item Special Toeplitz matrix. $\mathbf{B}$ is always a positive definite matrix, if it takes the following form:
    \begin{equation}
        \mathbf{B}=
\left[
  \begin{array}{cccc}
    1 & \beta & \cdots & \beta \\
    \beta & 1 & \cdots & \beta \\
    \vdots & \vdots &  & \vdots \\
    \beta & \beta & \cdots & 1 \\
  \end{array}
\right],
    \end{equation}
    where $0\leq \beta \leq 1$. In this case, the inter-cell interference between base stations is assumed to have the same value, i.e., $\beta_{i,j}=\beta$ for all $i\neq j$.
    \item Diagonally dominant matrix. The matrix $\mathbf{B}$ is said to be diagonally dominant if, for every row of the matrix, the magnitude of the diagonal entry in a row is greater than or equal to the sum of the magnitudes of all the other (off-diagonal) entries in that row, i.e.,
    \begin{equation}
        1=\beta_{i,i}\geq \sum_{j\neq i} \beta_{i,j},~ \forall i.
    \end{equation}
    In this case, it can be understood that intra-cell interference is more significant than inter-cell interference.
    This scenario may occur in heterogeneous cellular networks, which include some combination of macrocells, picocells, and femtocells \cite{mirahmadi2014interference}.
\end{enumerate}

\subsection{Performance Analysis of ETSC-MM Algorithm}
In this subsection, we present numerical results on applying the proposed algorithms to design unimodular/non-unimodular sequences with low ETSC.

\begin{figure}[ht]
  \centering
  \includegraphics[width=8cm]{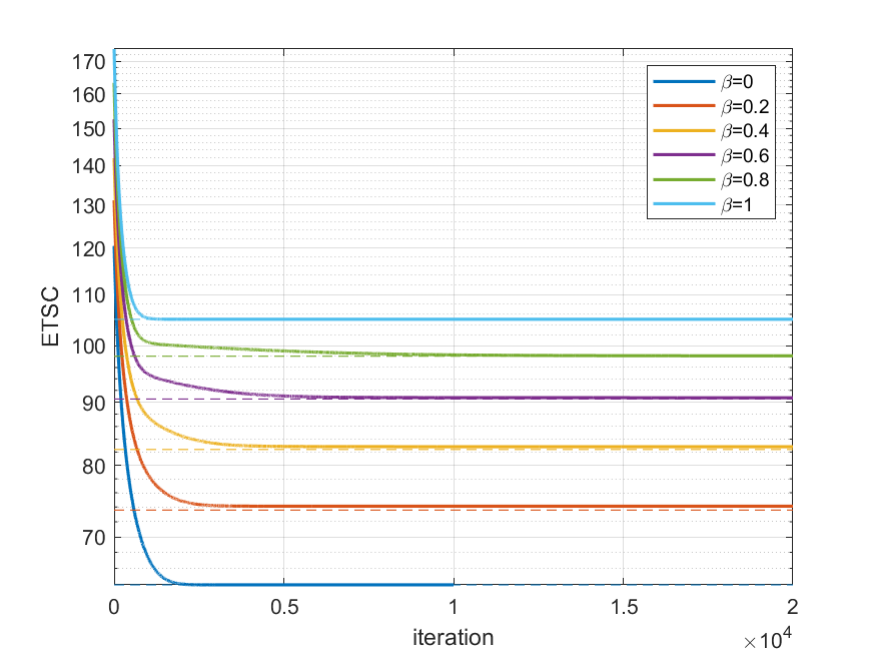}
  \caption{ETSC of non-unimodular sequence sets with $\tau=39$, $K=32$, $J=2$ and different $\mathbf{B}$.}\label{fig_39_32_2_C}
\end{figure}

\begin{figure}[ht]
  \centering
  \includegraphics[width=8cm]{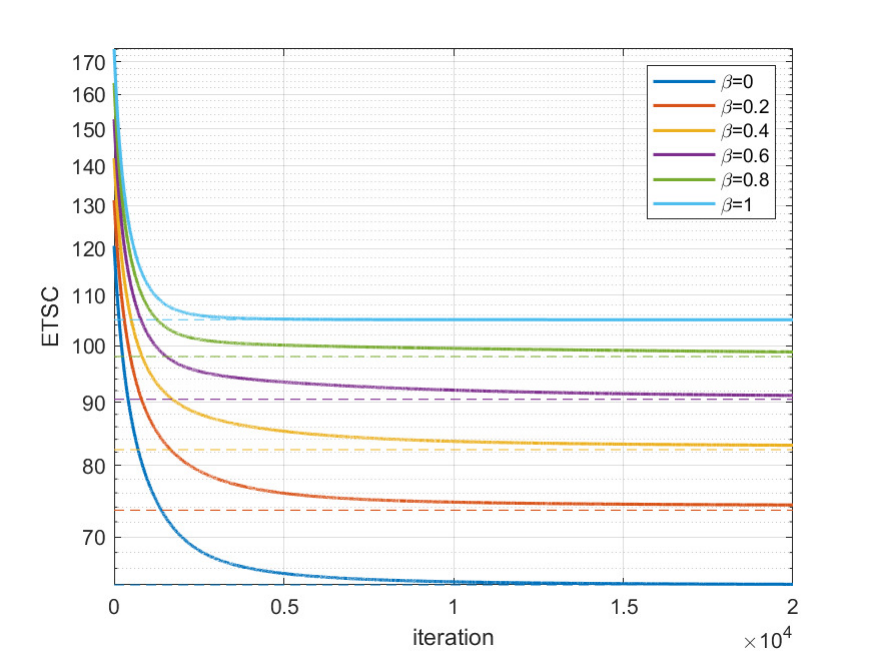}
  \caption{ETSC of unimodular sequence sets with $\tau=39$, $K=32$, $J=2$ and different $\mathbf{B}$.}\label{fig_39_32_2_U}
\end{figure}

In the first experiment, we compare the quality, measured by the ETSC defined in (\ref{equation_ETSC}) in the case $J=2$ (recall that the lower the better).
Using ETSC-MM algorithm, we have designed 6 sequence sets of parameters $\tau=39, K=32$ with different
$\mathbf{B}=
\left[
  \begin{array}{cc}
    1 & \beta \\
    \beta & 1 \\
  \end{array}
\right],\beta=0,0.2,0.4,\ldots,1.$
Figs. \ref{fig_39_32_2_C} and \ref{fig_39_32_2_U} show the objective function curves of the ETSC-MM algorithm for generating non-unimodular and unimodular sequences, respectively.
Here, the same random sequences are used as the initial seed sequence, and the stopping criterion is set to a maximum number of iterations of $2\times 10^4$.
The dashed line in the figures represents the extended Welch bound \cite{wang2020extending}. Figs. \ref{fig_39_32_2_C} and \ref{fig_39_32_2_U} show that the sequence sets generated by ETSC-MM algorithm almost meet the extended Welch bound for different $\mathbf{B}$.

\begin{figure}[ht]
  \centering
  \includegraphics[width=8cm]{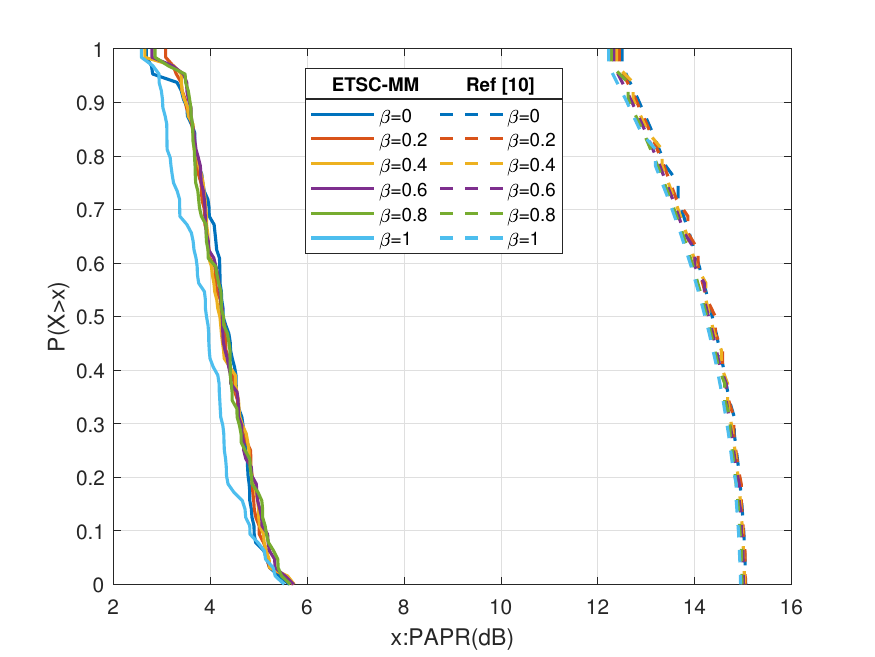}
  \caption{Comparison of PAPR with $\tau=39$, $K=32$, $J=2$ and different $\mathbf{B}$.}\label{fig_PAPR}
\end{figure}

Fig. \ref{fig_PAPR} compares the Complementary Cumulative Distribution Function (CCDF) of the PAPR for non-unimodular sequences generated by the ETSC-MM algorithm and the algorithm proposed in \cite{wang2020extending}.
It shows that the PAPR of the sequences generated by the ETSC-MM algorithm is predominantly below 6 dB. Furthermore, it is important to note that the ETSC-MM algorithm can also generate unimodular sequences, for which the PAPR is 0 dB.

\begin{figure}[ht]
  \centering
  \includegraphics[width=8cm]{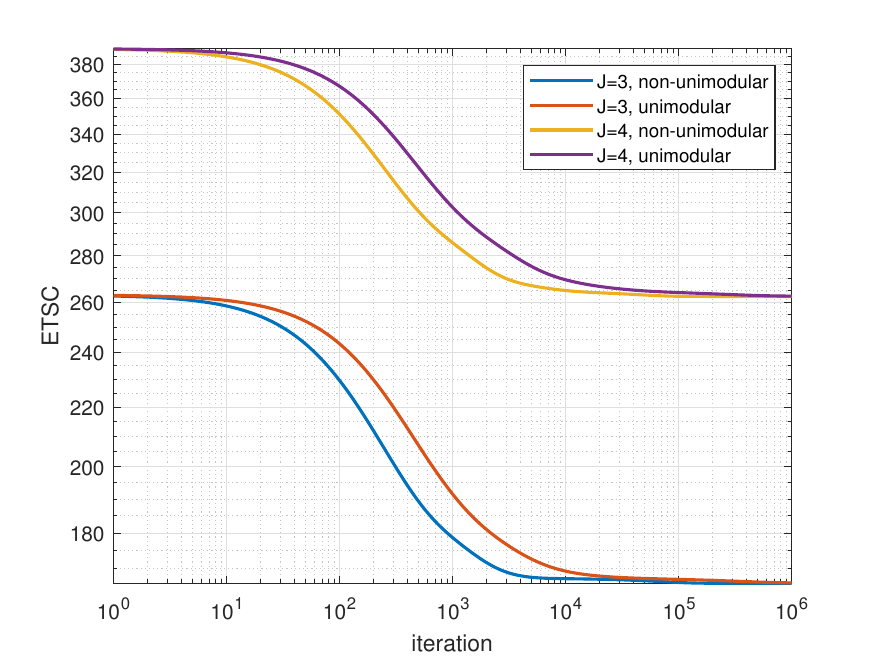}
  \caption{ETSC of sequence sets with $\tau=39$, $K=32$ and $J=3,4$.}\label{fig_39_32_multicells}
\end{figure}

In the second experiment, we compare the quality, measured by the ETSC defined in (\ref{equation_ETSC}) in the case $J\geq3$.
Using ETSC-MM algorithm, we have designed 4 sequence sets of parameters $\tau = 39, K = 32$ and
\begin{align*}
\mathbf{B}=&
\left[
  \begin{array}{ccc}
    1 & 0.8 & 0.2 \\
    0.8 & 1 & 0.6 \\
    0.2 & 0.6 & 1 \\
  \end{array}
\right], \hbox{ when } J=3, \\
\mathbf{B}=&
\left[
  \begin{array}{cccc}
    1 & 0.8 & 0.5 & 0.2 \\
    0.8 & 1 & 0.4 & 0.3 \\
    0.5 & 0.4 & 1 & 0.7 \\
    0.2 & 0.3 & 0.7 & 1 \\
  \end{array}
\right], \hbox{ when } J=4.
\end{align*}
Fig. \ref{fig_39_32_multicells} shows the objective function curves of the ETSC-MM algorithm for generating non-unimodular and unimodular sequences.
It shows that the ETSC-MM algorithm exhibits good convergence and monotonicity when generating sequence sets for multiple cells ($J\geq 3$).
In addition, although the objective function decreases more slowly when generating unimodular sequences, their performance is almost identical when the number of iterations is sufficiently large.
Please note that \cite{wang2020extending} cannot generate sequence sets for multiple cells.

\begin{figure}[ht]
  \centering
  \includegraphics[width=8cm]{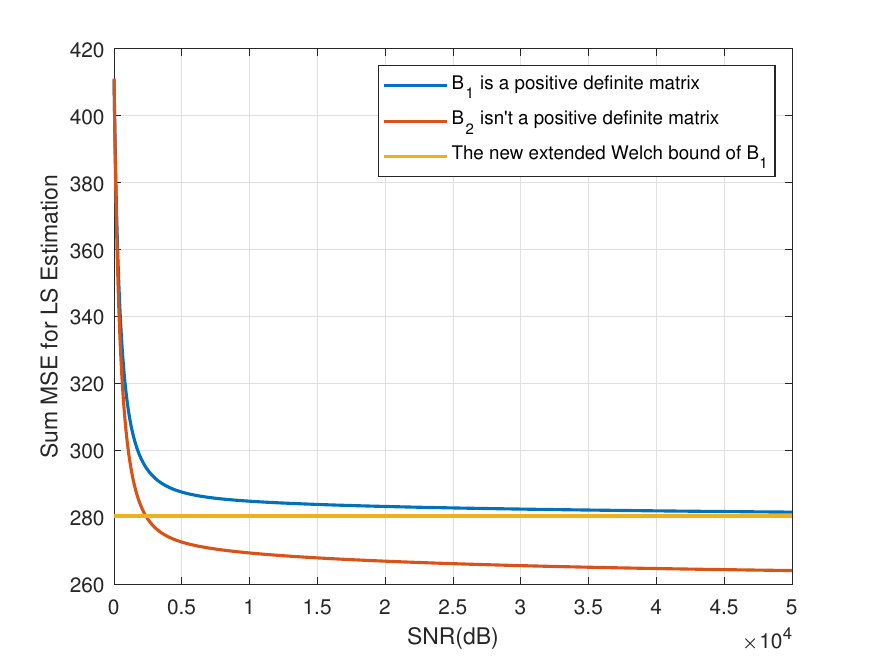}
  \caption{ETSC of sequence sets with $\tau=39,K=32,J=3$ and different $\mathbf{B}_1,\mathbf{B}_2$.}\label{fig_B1B2}
\end{figure}

In the third experiment, we compare the quality, measured by the ETSC defined in (\ref{equation_ETSC}) in the case $\tau = 39, K = 42, J=3$.
Using ETSC-MM algorithm, we have designed 2 unimodular sequence sets with different matrices $\mathbf{B}$:
\begin{align*}
\mathbf{B}_1=&
\left[
  \begin{array}{ccc}
    1 & 0.8 & 0.2 \\
    0.8 & 1 & 0.6 \\
    0.2 & 0.6 & 1 \\
  \end{array}
\right], \hbox{ it is a positive definite matrix} \\
\mathbf{B}_2=&
\left[
  \begin{array}{ccc}
    1 & 1 & 0 \\
    1 & 1 & 0.6 \\
    0 & 0.6 & 1 \\
  \end{array}
\right], \hbox{ it is not a positive definite matrix.}
\end{align*}
From Fig. \ref{fig_B1B2}, it can be observed that when $\mathbf{B}_1$ is a positive definite matrix, the ETSC-MM algorithm almost achieves the theoretical bound. When $\mathbf{B}_2$ is not a positive definite matrix, the ETSC-MM algorithm can find sequence sets with lower ETSC.

\begin{figure}[ht]
  \centering
  \includegraphics[width=8cm]{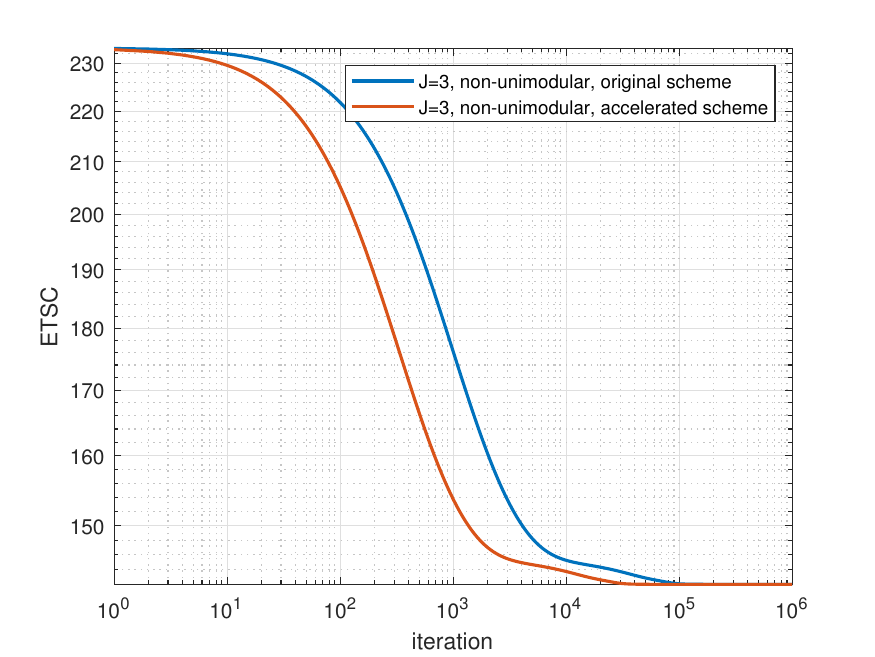}
  \caption{The comparison between the accelerated scheme and the original scheme for parameters $\tau=39$, $K=32$ and $J=3$.}\label{fig_39_32_OvsA}
\end{figure}

In the fourth experiment, we show the performance of the accelerated scheme compared to the original MM scheme.
Here, the sequence set parameters are the same as those in the second experiment.
Fig. \ref{fig_39_32_OvsA} shows the objective function curves for the original scheme and the accelerated scheme when generating non-unimodular sequence sets. It shows that the accelerated scheme effectively reduces the number of iterations. Note that the accelerated scheme was used in both the first and second experiments.

\begin{figure}[ht]
  \centering
  \includegraphics[width=8cm]{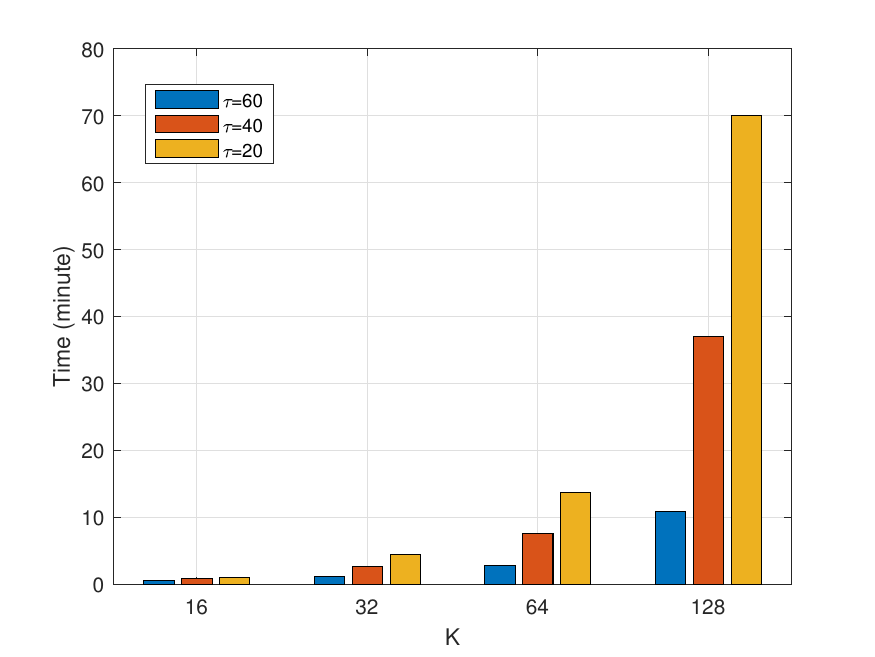}
  \caption{Comparison of the running time of the ETSC-MM algorithm.}\label{fig_TimeCeshi}
\end{figure}

The final experiment shows the actual runtime of the ETSC-MM algorithm on the Matlab platform, which serves as an intuitive illustration of the computational complexity analysis presented in subsection IV.C.
Fig. \ref{fig_TimeCeshi} shows the runtime for $10^4$ iterations with $N=JK=32,64,128,256$ and $\tau=20,40,60$.
It can be observed that as $\tau$ increases, the runtime exhibits a linear growth, while as $N$ increases, the runtime demonstrates a superlinear growth. This aligns closely with the theoretical analysis presented in subsection IV.C.

\subsection{Simulation Results}
In this subsection, we evaluate the pilot sequences generated by ETSC-MM and the theoretical construction.

\begin{figure}[ht]
  \centering
  \includegraphics[width=8cm]{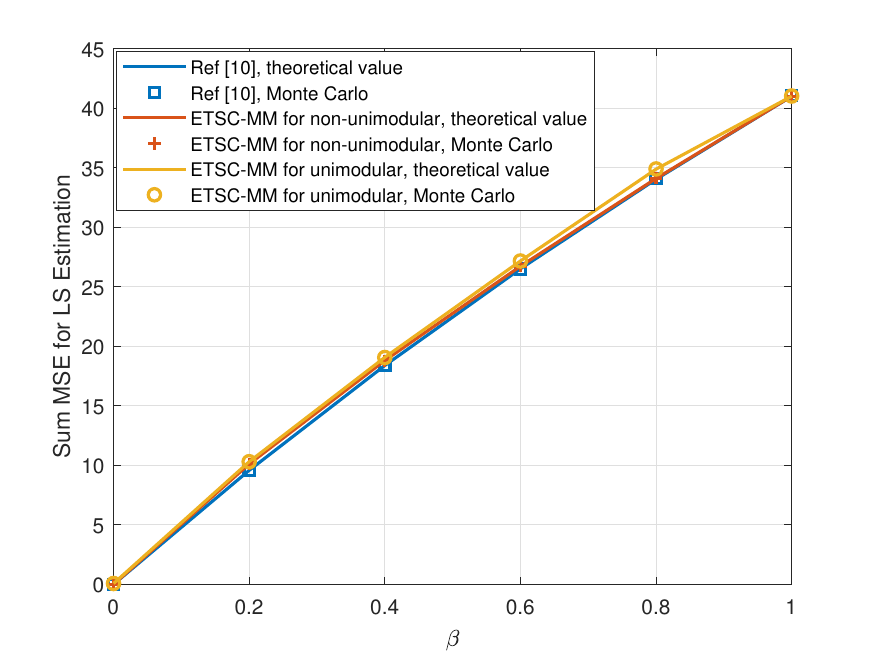}
  \caption{Sum MSE comparison with parameters $\tau=39$, $K=32$, $J=2$ and different $\mathbf{B}$.}\label{fig_Sim_39_32_2}
\end{figure}

In the first simulation, we compare the channel estimation performance of the sequence set generated by \cite{wang2020extending}, and the sequence set generated by the ETSC-MM algorithm for the case of $J=2$ and $K\leq\tau$.
Note that under these parameters, the sequences generated by [10] can meet the previous extended Welch bound, which can be considered as a theoretical limit. Fig. \ref{fig_Sim_39_32_2} shows the channel estimation performance of the unimodular/non-unimodular sequences generated by the ETSC-MM algorithm with all parameters being the same as in the second experiment.
From Fig. \ref{fig_Sim_39_32_2}, it can be observed that the three curves almost coincide.
Therefore, the ETSC-MM algorithm demonstrates almost optimal performance under these parameters.

\begin{figure}[ht]
  \centering
  \includegraphics[width=8cm]{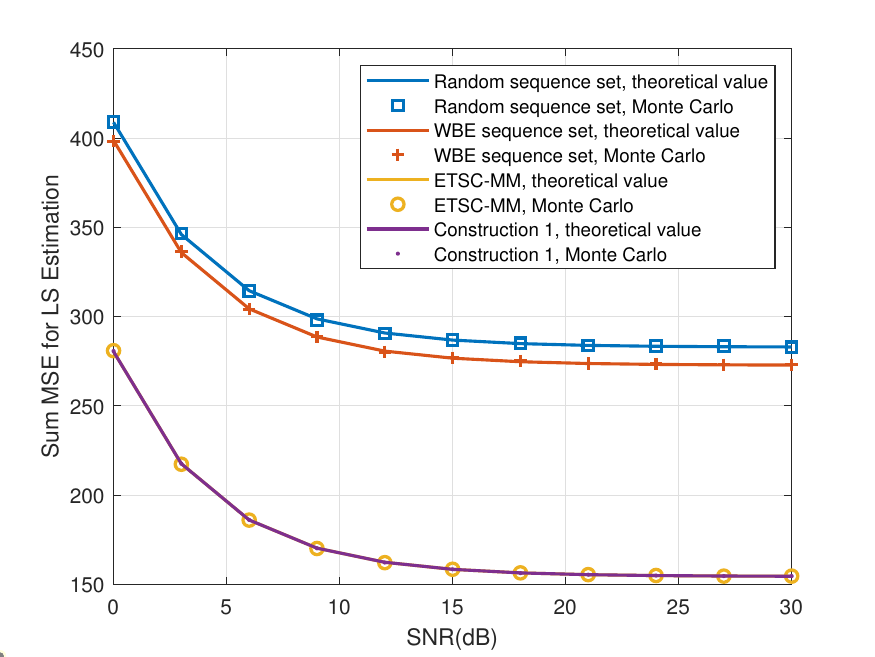}
  \caption{Sum MSE comparison with parameters $\tau=39$, $K=42$, $J=3$}\label{fig_Sim_39_42_3}
\end{figure}

In the second simulation, we compare the channel estimation performance of the WBE sequence set, random sequence set and the sequence set generated by the theoretical construction and the ETSC-MM algorithm for the case of $J=3$ and $K\geq\tau$.
Note that under these parameters, there are no reported methods in the literature for generating pilot sequence sets for such systems. Therefore, we use random sequences and WBE sequence sets for comparison. Fig. \ref{fig_Sim_39_42_3} shows the channel estimation performance of these sequence sets with parameters $\tau=39$, $K=42$, $J=3$, and $\mathbf{B}$ being the same as in the second experiment.
From Fig. \ref{fig_Sim_39_42_3}, it can be observed that the sequences generated by the ETSC-MM algorithm and \textit{Theorem} \ref{Theorem_NEWB} both exhibit optimal performance, significantly outperforming both the random sequences and the WBE sequence sets.

\begin{figure}[ht]
  \centering
  \includegraphics[width=8cm]{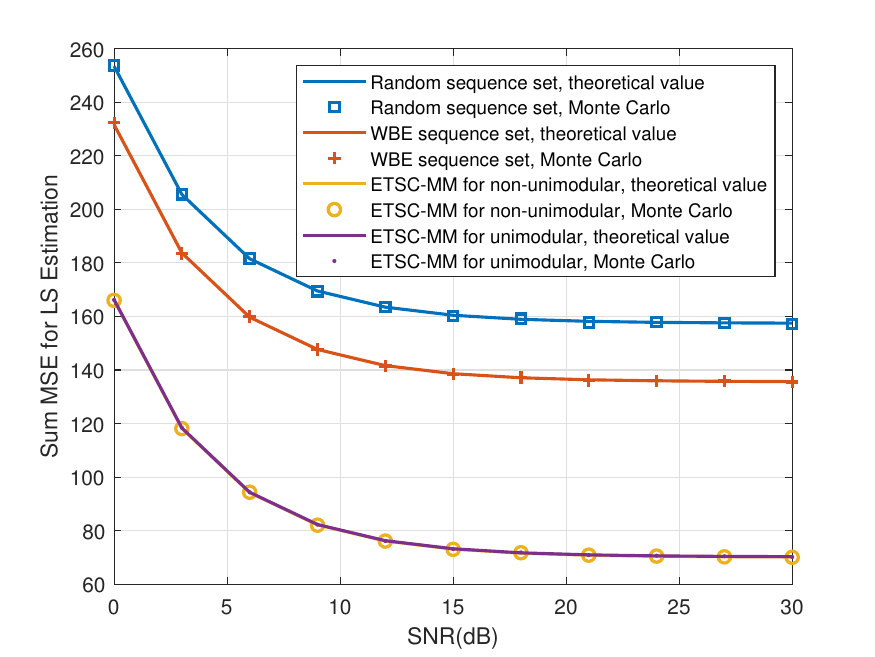}
  \caption{Sum MSE comparison with parameters $\tau=39$, $K=32$, $J=3$.}\label{fig_Sim_39_32_3}
\end{figure}

In the third simulation, we compare the channel estimation performance of the WBE sequence set, random sequence set and the unimodular/non-unimodular sequence set generated by the ETSC-MM algorithm for the case of $J=3$ and $K\leq\tau$.
Note that, as in the second simulation, there are no reported methods in the literature for generating pilot sequence sets for such cases. Furthermore, \textit{Theorem} \ref{Theorem_NEWB} cannot generate such sequence sets. Therefore, we use random sequences and WBE sequence sets for comparison.
Fig. \ref{fig_Sim_39_32_3} shows the channel estimation performance of these sequence sets with parameters $\tau=39$, $K=32$, $J=3$, and $\mathbf{B}$ being the same as in the second experiment.
From Fig. \ref{fig_Sim_39_32_3}, it can be observed that the unimodular and non-unimodular sequence sets generated by the ETSC-MM algorithm exhibit similar performance, significantly outperforming both the random sequences and the WBE sequence sets.

\section{Conclusion}
In this paper, we proposed the Cross-inner Product Theorem, which can be viewed as a generalized form of the previous Inner Product Theorem. And we provided a new extended Welch bound based on the Cross-inner Product Theorem, which offers a theoretical lower bound on ETSC for sequence sets with $J \geq 2$, $K \geq \tau$ and $\mathbf{B}$ is positive definite matrix.
Additionally, from the necessary conditions of the bound, the optimal sequence set can be easily obtained when the interference power factor matrix is positive definite.
For the case $K<\tau$ or $\mathbf{B}$ is not positive definite matrix, we proposed the ETSC-MM algorithm to generate sequence sets with low ETSC based on the MM framework.
Compared to the literature, the ETSC-MM algorithm not only generates unimodular sequences but also addresses the sequence generation problem for $J \geq 3$.
Numerical experiments show that the sequences generated by the ETSC-MM algorithm almost meet the theoretical bound (if such the corresponding bound exists).
To more clearly present the above contributions of this paper, the research relationships are shown in TABLE \ref{table1}.

\begin{table*}[ht]
\centering
\caption{Existing and proposed methods and bounds}\label{table1}
\begin{tabular}{|c|c|c|c|}
  \hline
  Ref & {Constraints} & Construction methods & Theoretical bound \\ \hline
  \cite{welch1974lower} & $JK\leq \tau$ and $J,K,\tau \in \mathbb{N}^+$ & Theoretical construction based on orthogonal matrix & 0 \\
  \hline
  \cite{wang2020extending} & $K\leq \tau \leq JK$ and $K,\tau\in \mathbb{N}^+,J=2$ & Numerical algorithm & $\frac{2K^2(1+\beta)}{K+\beta(\tau-K)}$ \\
  \hline
  Theorem 2 & $\tau\leq K$, $J,K,\tau \in \mathbb{N}^+$ and $\mathbf{B}$ is positive matrix & Theoretical construction based on WBE sequence & $\frac{K^2}{\tau} \sum_{i=0}^{J-1} \sum_{j=0}^{J-1} \beta_{i,j}$ \\
  \hline
  Algorithm 1 & $J,K,\tau \in \mathbb{N}^+$ & Numerical algorithm & Unknown \\

  \hline
\end{tabular}
\end{table*}
\bibliographystyle{IEEEtran}
\bibliography{Refsnew}

\end{document}